# Finding Alternate Paths in the Internet:
# A Survey of Techniques for End-to-End Path Discovery

SAMEER QAZI AND TIM MOORS

*PN Engg College, National University of Sciences and Technology, Islamabad, Pakistan*

*University of New South Wales, NSW, (Sydney), Australia.*

Abstract: The Internet provides physical path diversity between a large number of hosts, making it possible for networks to use alternative paths when one path fails to deliver the required Quality of Service. However, for various reasons, many established protocols (e.g. de facto Internet inter-domain routing protocol, Border-Gateway Protocol - BGP) do not fully exploit such alternate paths. This paper surveys research into techniques for discovering end-to-end alternate paths, including those based on monitoring path performance, choosing paths that are maximally disjoint, and in routing across multiple paths. It surveys proposals for making BGP better able to exploit multiple paths and how multi-homing can create alternate paths. It also describes how alternate paths can be realized through detour routing (application layer mechanisms) and routing deflections (network layer mechanisms). It also discusses Fast Re-Route techniques for construction of backup routes. It concludes by surveying open research issues into the discovery and use of alternate paths in the Internet.

INTRODUCTION

The Internet has expanded to a massive scale, incorporating millions of devices belonging to tens of thousands of networks [1]. One feature that has enabled this scaling has been its use of hierarchical routing, in which separately administrated Autonomous Systems (*ASes*) can independently choose their own interior routing protocol (e.g. *OSPF* or *IGRP*) and are interconnected by a single exterior routing protocol, the Border Gateway Protocol (*BGP*). Whereas interior routing protocols can choose paths based on performance metrics selected by the administrator, *BGP* ignores such performance metrics, and only considers routing policies in trying to find a route. This design of *BGP* is partially a response to the difficulty of reaching consensus across all *ASes* as to what performance metrics should be used and optimized, partly because merely accounting for service provider policies is sufficiently challenging in itself, and partly because link and device performance are dynamic, and accounting for their variations would limit the scalability of *BGP*. Consequently, routes across the Internet are often not optimized for performance. Yet many applications are sensitive to route performance. At one extreme, a route that simply fails to deliver packets will clearly impinge on applications that communicate across that route. *BGP* will eventually detect and recover from such faults, but to permit it to scale, *BGP* does not frequently disseminate path availability information, e.g. it may sometimes take several minutes to learn and apply path updates [2]. As a result, applications may experience lengthy network outages. A less extreme example of sensitivity to route performance is that of real-time applications such as Voice over IP (*VoIP*) that are sensitive to the delay with which information is transferred across the network. For these applications, the connectivity that *BGP* provides may be insufficient, since they seek a certain Quality of Service (QoS) from end-to-end routes that they use.

Several independent research findings [3-4] have previously shown evidence of path diversity in the Internet. This paper focuses on finding such alternate paths so as to improve end-to-end QoS as outlined by Figure 1(a). Savage et al. [4] showed that for almost 80% of the paths used in the Internet there exists an alternative route with a lower probability of packet loss, and that for 15% of the paths there exists an alternative that improves latency by more than 25%.

A few examples of where an alternate route between source and destination can benefit applications are highlighted in Figure 1. For many of the user perceived performance failures/faults, e.g. delay in loading a web page or patchy audio in a *VoIP* session, there may exist a less congested path between source and destination (Figure 1a). An alternate path may also exist by virtue of content being replicated at many mirror sites across the Internet (Figure 1b). A more questionable application of an alternate path may be to connect two hosts if the routing administration of one (e.g. as enforced by a firewall) blocks incoming connections from the other due to policy/security reasons (Figure 1c). In such cases, a composite alternate path can be formed by first directing packets towards a third (intermediate) host which is not blocked by the firewall. Such collaboration of end hosts can also be used to improve QoS; by directing packets towards a third host to detour around a fault on the primary Internet path. Such networks are often termed as *RONs* -Resilient Overlay Networks [5] and are discussed in more detail in Section IV. The composite alternate path through a third host is called an *overlay path* or more specifically a *one-hop overlay path*. Similarly, reliability of transmission under large link loss rates may be improved by simultaneously sending redundant packets over multiple alternate paths (Figure 1d).

Having introduced some of the uses of alternate paths, this paper will now survey techniques for discovering such paths. Section II starts by examining what features are desirable in alternate paths, and thus the criteria for selecting alternate paths. Section III considers legacy approaches to using multiple paths, through extending BGP and through the use of multi-homing. Section IV considers how alternate paths can be used for such purposes as detour routing, routing deflections and backup routes. Section V surveys some of the open research issues relating to discovering and using alternate paths, and Section VI concludes the paper and proposes three specific areas that are particularly worthy of future research.

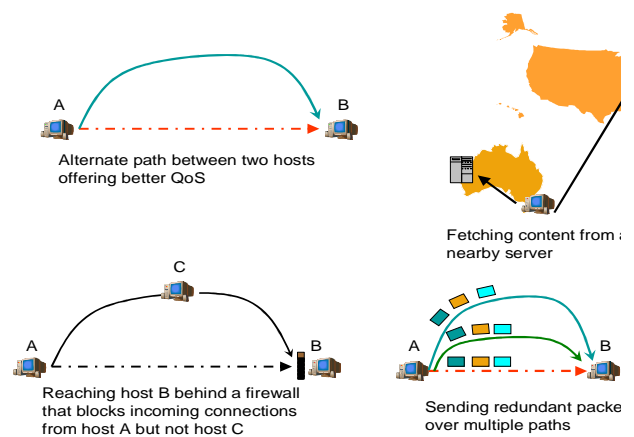

Figure 1. How alternate paths can be useful. (a) (top-left) an alternate path (solid) offering better QoS than the direct path (dashed); (b) (top-right) alternate path by virtue of content replication on a nearby server; (c) (bottom-left) Reaching a host behind a firewall using a composite alternate path through a third host; and (d) (bottom-right) sending content over multiple alternate paths.

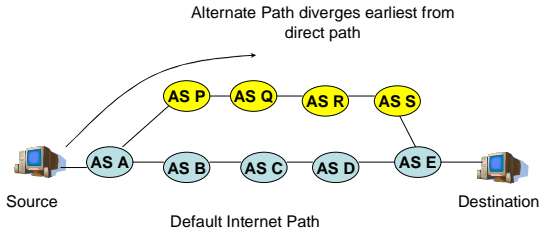

Figure 2. Earliest-Divergence Heuristic [10] to select disjoint alternate paths

CRITERIA FOR SELECTING ALTERNATE PATH/S

One criterion for judging alternate paths is in terms of their performance benefits over the primary Internet path (Section II.1), e.g. in terms of such metrics as latency, throughput and packet loss. Another criterion for selecting alternate paths is to select the most disjoint paths (Section II.2) in the hope that a failure on the primary Internet path will not affect the alternate path. When all paths are prone to failures, the mere existence of alternate paths is important for sending redundant data on multiple paths (Section II.3) to facilitate timely data delivery for crucial Internet applications, e.g. sending video/audio streams.

*Monitoring Paths Based on Performance*

Dynamic path monitoring is essential in order to quickly recover from a failure in the underlay network by using an alternate path. A large body of research discusses choices of an appropriate performance metric such as latency, throughput and loss rates for selecting backup paths. Paths are ranked on the basis of these metrics using *scoring functions*; these range from weighted-moving averages over finite temporal windows to statistical approaches [5-7].

*RON* [5] distinguished paths on the basis of latency, throughput and loss rates, making the ranking of paths application-specific. Zhu [7] used *available-bandwidth* for alternative path selection, claiming that latency, loss rates and throughput metrics could be 'misleading' as they often depend on the protocol implementations, network heterogeneity or temporal effects. Zhu argues that throughput is a function of TCP parameters and that thresholds set for detecting allowable loss rate and latency could be misleading because of the dynamism and heterogeneity experienced by the network. Similarly, Lee et al. [8] measured capacity between paths and selected paths based on *available bandwidth* criteria. Hu and Steenkiste [9] showed that in comparison to delay and loss rates, estimation of bandwidth is relatively easy since it is often bounded by the bandwidth of bottleneck links. Identification of such bottleneck links is often easy since as the links in the core of the Internet are often over provisioned, so bottlenecks often appear within three to four IP hops of the end hosts that are monitoring the paths in order to create an overlay network.

*Disjoint Paths*

Several researchers [10-12] argued that since Internet paths are often stable on time-scales of days, maintaining complete physical topology information about default and alternate Internet paths allows one to select the most disjoint alternate path without having to continuously monitor path performance. Fei et al [10] showed that an *Earliest Divergence* Rule (Figure 2) can work well by selecting, as the alternate path, the path which diverges from the default-path at the earliest point near the source. Qazi and Moors [11] similarly investigated a *Maximum Divergence* Rule to pick an alternate path that is most divergent from both the source and the destination parts of the original path. However, these techniques assume availability of detailed information about which Autonomous Systems ($ASes$) the primary and the alternate paths pass through. Such knowledge is sometimes difficult to obtain. Traceroutes and other tools used for mapping paths are known to reveal path information inaccurately under certain conditions [13-19].

Selecting alternate paths based on disjointness may help recover from path outages, but is sometimes inefficient in ensuring strict application-specific metrics, like delay, throughput etc. For example path delays may not always be a simple function of fiber delays but a combination of fiber delays, congestion on individual links and packet queuing delays in routers. This makes path monitoring to meet application-specific QoS demands more important than merely ensuring spatial diversity. Nevertheless, the bulk of the thrust of new research is centered on improving design heuristics to choose disjoint paths.

Instead of using dynamic online algorithms to monitor and select alternate paths, offline processing of path measurements can be used to reveal spatial relationships (disjointness) between paths. Cui et al. [6] proposed a unique method which establishes performance-related correlations among the behavior of paths e.g. link-latency. Such metrics can then be used to find a backup-path for a given primary-path between two hosts, with least correlated-failure probability.

*Multi-path routing*

When paths are more failure prone, path performance varies widely over short periods of time, so for longer duration data transfers; e.g. video streams, it is infeasible to select *one* alternate path for QoS optimization. Under such circumstances, redundant data may be sent over multiple paths, in the hope that at least one copy of each data packet will be received correctly. Research [5] indicates that alternate paths between end hosts may fail independently of each other, since routing domains which are independently administered rarely share underlay links. To further reduce the probability of packet loss, advanced encoding schemes, e.g. Forward

Error Correction ($FEC$), may be used to detect and correct errors, and hence tolerate packet losses.

Antonova et al. [20] investigated the optimal breakdown of traffic across multiple paths when sending a video stream with bounded delay requirements.

While Zhao [21] claimed positive results of using constrained multi-cast to ensure end-to-end connectivity in the face of failures, Anderson et al. [22] concluded that such schemes are only useful when links suffer from low levels of congestion. Moreover, another alarming finding by the same study is the fact that failures on alternate paths may be more correlated than previously imagined: a packet loss on one path increases the conditional loss probability of the redundant packet on an alternate path by about 60 percent. Even packet-encoding schemes such as $FEC$ lose their effectiveness when path failures are correlated. Moreover, a large number of packets sent on the network unnecessarily consume network resources, increase network load and rob other flows of their fair share of network resources.

LEGACY AND MODERN APPROACHES FOR IMPROVING BGP TO FACILITATE ALTERNATE PATH DISCOVERY

The Border-Gateway Protocol ($BGP$) is the de facto Internet inter-domain routing protocol connecting all networks into one giant Internet. $BGP$ in many instances logs several policy-oriented paths to a single destination. For example, Figure 3 shows paths to the destination prefix 213.145.13.0/24, including paths to all destinations with $IP$ addresses ranging from 213.145.13.0 to 213.145.13.255. There are three possible paths; of these the path $B-C-D$ in chosen in accordance with $BGP$ preference. This path will continue to be used even if it offers suboptimal performance, provided the destination is still reachable. Now suppose that the inter-AS link between ASs $C$ and $D$ breaks, rendering the path $B-C-D$ inoperable. Only when the $BGP$ speaker in AS $C$ realizes that its connection with AS $D$ has broken will messages about this fault spread in the network. AS $C$ will inform its neighboring $BGP$ border router in AS $B$ that the destination prefix 213.145.13.0/24 is no longer reachable through it. The bordering $BGP$ speaker in AS $B$ will then convey this to the border $BGP$ router in AS $A$, which will then select the next best route according to policy. Note that the distributed nature of such message exchanges is often much more complicated and time consuming than the toy example just presented. As a result, alternative routes offering better QoS remain unexploited; due to the scalability objectives of $BGP$.

New extensions for $BGP$ have been proposed to alleviate some of the problems of delayed path convergence. For example, in the previous example mentioned, AS $C$ only indicates that it is unable to reach the destination prefix through AS $D$ but does not specify the cause of this fault. As a result the next best routes selected by $BGP$ do not take into account the cause of the fault, so these next best routes may themselves have failed before $BGP$ eventually converges on routes that are not affected by the failure. Delayed path convergence of $BGP$ has been effectively addressed through simple techniques such as flooding the network with the cause of a path failure to quickly rid the network of stale routes [23] (Figure 4). John et al. [24] proposed *Consensus Routing* to improve the consistency of inter-domain routes in $BGP$ by separating the functions of packet forwarding and route computation.

Kushman et al. [25] proposed an architecture $R-BGP$ whereby alternative disjoint fail-over routes are also announced by $BGP$ which enable quick failover (when possible) and guaranteed $BGP$ convergence without any routing loops. They provide detailed insight into this problem and explain which failover routes are appropriate to be announced and where in the AS hierarchy they should be announced.

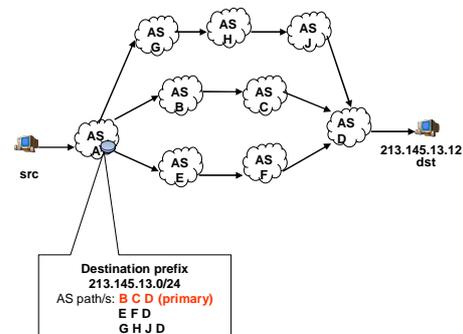

Figure 3. Paths to a typical destination prefix in a BGP table inside a BGP router.

Similarly, Quoitin et al. [26] proposed that several of the $BGP$ inter-domain static path selection parameters could actually be used for traffic engineering purposes, e.g. to force selection of better alternative paths. This could be achieved by selectively advertising destinations on different paths based on *IP* prefixes, artificially inflating the cost on one of the paths ( *AS* path-prepending) to discourage it from being selected, or by advertising a preference for a path to a neighboring *AS* explicitly through a *MED* (multi-exit discriminator) attribute. Similarly, the Local-Preference attribute that *BGP* uses to assign fixed weights to paths through dissimilar inter-domain bandwidth links could be made more sensitive to dynamic performance through active path measurements. Another technique through which an *AS* can exploit inter-domain path diversity is to tweak its own Interior Gateway Protocol (*IGP*), which is used to select an inter-domain path that leads to least internal (intra-domain) cost. This could end up constantly selecting one of several egress points towards other *ASes*. More granular *BGP* weight tuning could exploit path diversity to choose other paths.

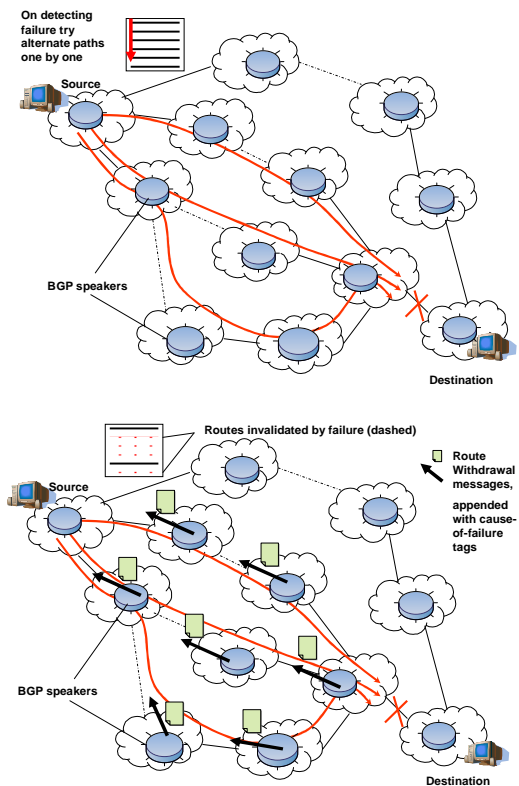

Figure 4a. (top) A single link-failure invalidates several valid routes (shown by bold-red arrows); BGP will select the next best path based on policy. Figure 5b. Appending path-withdrawal messages with 'cause-of-failure' tags [23] help eliminate all invalid routes quickly and converge to valid route quickly

Multi-homing is another technique through which hosts at the edge, or transit providers in the core, of the Internet maintain redundant connections which can be exploited for the purposes of fault tolerance. Host (stub) domains may announce one or more connections to one or more ISPs over one or more IP addresses [27].

The early approach to multi-homing was quite liberal: Stub domains could acquire special Provider Independent (*PI*) addresses from their Regional Internet Registry (*RIR*). *PI* addresses are globally unique and are not assigned by transit providers for their assigned address blocks. For example, if a stub domain that is multi-homed to two provider networks is assigned a *PI* address of 20.0.120.0/24 (Figure 5a), than it can advertise this to both of its transit providers which will propagate it to their own upstream providers, where it will reach other parts of the Internet to provide dual connections for the host domain.

Using *PI* addresses was a simple approach to multi-homing. However, this led to scalability issues together with the problem of depleting the limited *IPv*4 address space. Presently, stub domains are only allowed to use Provider Aggregatable (*PA*) addresses, which are *IP* sub-blocks from the *IP* address space assigned to their primary provider domain. Stub domains thus consider one of their immediate provider networks to be their primary *ISP* and the remainder as secondaries. This address is then advertised to its secondary *ISPs*. Revisiting the previous example, the secondary *ISP* would separately advertise the *PA* address of the multi-homed site, in addition to its own (as it cannot be merged with its own aggregate). On the other hand, the address block advertised by the primary *ISP* address would be the larger sub block 20.0.0.0/8; the smaller more specific sub-block 20.0.120.0/24 will be dropped to avoid inflating the size of *BGP* tables (Figure 5b). Since the Internet uses longest prefix matching when routing to destinations, the secondary *ISP* will thus be used to connect to the stub network for inbound packets due to the more specific prefix announced by it. Thus, the redundant paths cannot be used simultaneously to meet Traffic Engineering (*TE*) objectives or to achieve quick failover as dictated by the stub domain, as this traditional approach to multi-homing will again depend on *BGP* reaction time to provide a failover path. Also, note that even using *PI* addresses introduces one additional routing entry per multi-homed host. Huston [28] and Bu et al. [29] noted that the number of *BGP* routing entries in the Internet increased by an order of magnitude between 1995 and 2005.

USES OF ALTERNATE PATHS

Previous literature [24] has categorized three distinct methods of realizing alternate paths in the Internet: *detour routing* [3, 5], *routing deflections* [30] and *back-up route* construction [31].

Detour routing (Section IV.A) works when the primary Internet path is deemed to have failed, in which case packets are deflected towards an intermediate node in the hope that this will provide a detour around the failure on the primary path. Note that the location of this fault is not known. The intermediate node then directs the packets towards the original destination.

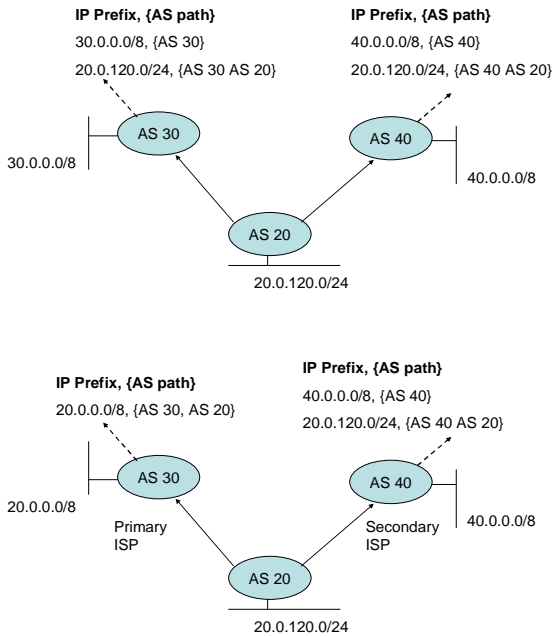

Figure 5. Multi-homing using (a) (top) Provider Independent (PI) and (b) Provider Aggregatable (PA) addresses

Routing deflections (Section IV.B) are different from detour routing in that deflection decisions are somewhat more localized. For example, if a router finds that the next hop link has failed it may forward the packet on an alternate link.

Back-up route construction (Section IV.C) is a more sophisticated approach in which a path between two end points is specially constructed to meet specific QoS requirements. Note that this path may differ from the path selected by the network itself. Back-up paths may also be selected (in addition to a primary one) based on maximum disjointness from the primary path, to avoid all possible failure scenarios. Such back-up paths allow quick failover once the primary path fails and are often also referred to as Fast Re-Route ($FRR$) construction.

*Detour Routing*

Gummadi et al. [3] showed that a large number of path and performance failures could be masked by detouring packets to an intended destination via an intermediate host located in an *AS* that is off the primary Internet path. Resilient Overlay Networks provide a systemic framework for exploiting the path diversity in the Internet based on this observation. RONs typically consist of a group of end hosts or network layer devices, e.g. routers, in the Internet that agree to route packets between each other through tunneling mechanisms to exploit the path redundancy in the Internet. Figure 6 shows the path between an end host in University of New South Wales (*UNSW*), Sydney, Australia and a host, www.example.com, located in California, US. A university such as *UNSW* typically uses the services of bigger provider *ISP* such as *AARNET* (Australian Advanced Research and Educational Network) to connect to hosts in the continental US. Such providers often use the hot potato routing principle [32], to try to shift this traffic off their network quickly at the nearest inter-domain egress point to send it to its US based destination. In the case of *UNSW* and *AARNET*, traceroute shows that the original path uses an egress point of *AARNET* at Sydney that takes the packets to www.example.com via a router in Honolulu, Hawaii to an ingress point in Los Angeles in the US.

RONs can exploit Internet path redundancy by deflecting packets away from the original path if it suffers an outage. Now consider the situation, if the end host in *UNSW* and the host www.example.com formed part of an overlay network together with another host inside *CMU* (Carnegie Mellon University). If *CMU* were to be used as the intermediate relay host (assuming there was a fault on the default path via Honolulu, or this path had become congested due to a sudden surge in traffic) then the new path would use an *AARNET* egress point at Sydney, as before, but takes the packets to a different ingress point inside the US (Seattle instead of Los Angeles). By taking such detours, RONs can use alternate paths to mask underlying path failures.

*Routing Deflections*

Yang and Wetherall [30] proposed that instead of using strict *least cost* or *shortest path* rules, packet forwarding decisions in routers should be more flexible and should allow choices of multiple potential next hop candidates, which allows exploiting of the path diversity (Figure 7). They showed that *deflection* is possible while forwarding packets at routers by selecting one of candidate choices. Moreover they showed that such deflections can select loop-free shorter paths without violating *ISP* rules. Routers only need to consider a few simple deflection rules while forwarding packets. This technique requires packets to be encoded by a shim-header (in between the network and transport header), which can potentially incur non-negligible packet processing overhead, in order to facilitate path deflection decisions. Moreover, these studies do not comprehensively explore the QoS benefits of these alternate paths. Also, such studies so far have only investigated the feasibility of the technique in a few large ISPs, e.g. Sprint and Abilene, where the range of path diversity might be exaggerated. Its practical benefits and ability to be deployed across the wider Internet are still uncertain when we consider that due to the power-law structure, there is a large degree of link sharing amongst paths [33-36], indicating that there may not be as many path deflection choices as the studies indicate. These questions are left as an open debate and hence need to be investigated thoroughly.

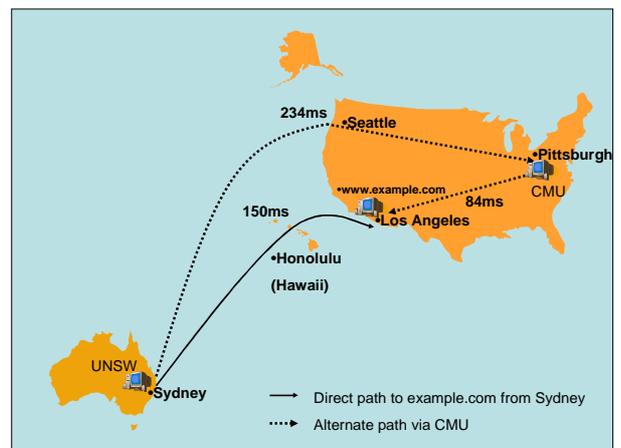

Figure 6. Direct path between UNSW and example.com and a one-hop overlay path via CMU.

Routing deflections can also be effectively implemented using more advanced multi-homing techniques. These include *Middle-box, Routing* and *Host Centric* approaches.

In the *Middle-box Approach*, a Network Address Translation-(*NAT*) box (or boxes) running Multi-homing Translation Protocol (*MHTP*) [37] or Multi-homing Aliasing Protocol (*MHAP*) [38] are installed at the edge of the multi-homed site. The main purpose of *NAT* boxes is to re-direct packets towards a working provider instead of relying on *BGP* to do this indirection after a failure.

The *Routing Based Approach* is based on enhancing *BGP* and *IGP* to support adequate multi-homing with *TE* objectives [39]. In this approach, a multi-homed site maintains two or more Site Exit Routers. The multi-homed site is provided with one *PA* address per provider. Route aggregation is achieved by announcing the appropriate aggregate to each provider. This solution is effectively single-homing to each provider rather than multi-homing.

In the *Host Centric Approach,* multi-homing relies on the ability of the host to detect path failures. For example if a host sees frequent packet loss on the path, it may change the source address on its packets so that the site exit router selects a different provider network for the outgoing packets. Such detection by the hosts can both be made using Transport layer and Network layer techniques [39].

### Back-up Routes

This section surveys how backup routes can be used to set up paths using specified QoS parameters, and techniques for fast reroute construction of inter-domain and intra-domain routes.

### Path set-up using specified QoS parameters

Multi-Path Inter-domain Routing (*MIRO*) [31] presented a comprehensive solution to address issues with *BGP* regarding QoS optimization of paths, proposing several architectural modifications to the current *BGP*. The architecture shows how it is possible for *ASes* to advertise multiple routes for destination-prefixes through on-demand path announcements known as *pull-based route retrieval.* Pull-based route retrieval consists of two main steps:

(i) a *route-negotiation* step, in which an interested *BGP* speaker floods a query for a route request that fulfills some criteria, and accepts routes advertised by peers and requested peers may return such paths through selective export policies so that other peers stay oblivious to this information exchange; and

(ii) *routing-tunnel establishment* where peers flood information amongst themselves for any successfully negotiated route (Figure 8).

This technique ensures that all such negotiated paths meet *BGP* policy constraints through selective export policies. Not only does the architecture meet all design objectives but it also proposes an evolutionary design-approach; offering attractive incentives to network-administrators adopting *MIRO* while at the same time making it possible for native-*BGP* users to co-exist.

### Fast Reroute Construction for Inter-domain and Intra-domain routes

Fast Re Route (*FRR*) considers construction of failover paths so that alternate back-up paths can be used immediately after detecting a failure, without having to wait to send/receive routing updates to/from neighboring routers. *FRR* can be used for both inter-domain routing, e.g. with *MPLS*, and *IP* -based intra-domain routing.

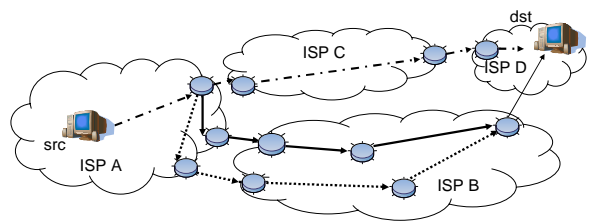

Figure 7. Path deflection decisions made at router level can exploit the path diversity in the underlay network

*MPLS-FRR*

*MPLS* (Multi-protocol Label Switching) can be used for inter-domain path set-up with intra-domain traffic engineering. Instead of switching (routing) packets at the network layer based on the inspection of destination addresses, the routes should be negotiated in the beginning according to the demands of the application. Once such a path has been found, the negotiated path segments and all packets belonging to the application are assigned specific labels, and routing takes place on the basis of these labels. Current efforts are dedicated towards improving its scalability and extending *MPLS* solutions to an inter-domain level.

There are two approaches [40] for inter-domain path set-up: Backward Recursive *PCE* (Path Computation Element)-based Path Computation (*BRPC*) and Per Domain Path Computation. In *BRPC*, the path is formed recursively from the destination towards the source domain. Every domain has certain "entry" boundary nodes (*BNs*) into the domain and "exit" *BNs* out of the domain (into other domains). Each domain (starting from the domain of the destination) constructs a Virtual Shortest Path Tree (*VSPT*) from the destination towards each entry *BN* of that domain. The *VSPT* is then sent to the *PCE* of the previous domain i.e. the one that is closer to the source domain (Figure 9), where it is concatenated to their Traffic Engineering Database (*TED*) (as links). This method of only exchanging *VSPTs* ensures confidentiality about the internal structure of the domains. This process is repeated until it reaches the source domain and subsequently the head end router of the *TE − LSP*.

In Per Domain Path Computation, each individual domain constructs a path segment between an entry and exit *BN*. The complete Label Switched Path (*LSP*) is then given by the concatenation of these computed path segments inside individual domains. If a domain is unable to find a suitable path, it incorporates a special crankback mechanism [40-42]. When one of the Next Hop (*NH*) domains (*ASes*) is unable to find such a path, they may refer a failure message to the preceding domain (*AS*). This message will then be conveyed to the *PCE* (of the preceding domain) which will re-compute path selection criteria so as to exploit different egress point/s to different *NH* domain/s (*ASes*).

To select a path conforming to the QoS requirement of the *LSP* request, the *PCE* uses *TEDs* maintained by *IGP* / *IS − IS* protocols with *TE* extensions. *PCE* may also return primary and backup *LSPs* for failover if requested. The primary novelty of *MPLS − TE* is in the three areas of: (a) extending the *FRR* concept to an inter-domain level; (b) its approach of considering more dynamic path properties than just exploiting path diversity and (c) computation of back-up *LSPs* when primary *LSPs* fail.

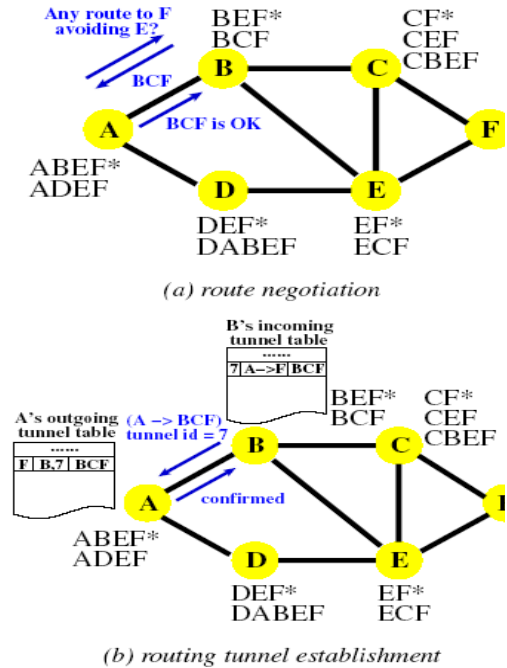

Figure 8. MIRO routing example[31]

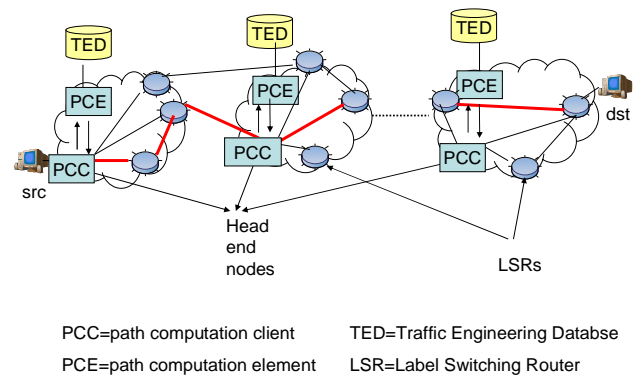

PCC=path computation client     TED=Traffic Engineering Databse
PCE=path computation element     LSR=Label Switching Router

Figure 9. Inter-domain MPLS path construction

*IP-FRR for IGP*

Link state protocols (e.g. *OSPF / IS − IS* ) used as *IGPs* (Interior Gateway Protocols) converge much faster than *BGP* (a path vector based protocol) owing to the small scale of interior networks. Recovery times of sub 200ms are not uncommon [43]. Such small delays often go unnoticed even by *VoIP* customers who demand quick failover. Interestingly, the majority of this time is not spent on detection of failure, flooding new routing information (updates) and re-computing routes, but in loading the revised forwarding tables into the router's Forwarding Information Base ( *FIB* ) [44]. Having pre-computed alternative path information available, which avoids failed components, can definitely help in quick recovery.

Failover paths inside a domain are considered so that individual routers can try alternate paths instead of waiting to send/receive routing updates to/from neighboring routers. For example, routers could identify Shared Risk Link Groups ( *SRLGs* ), i.e. a set of links that fail together owing to a physical commonality between them e.g. adjacency to the same router. Various proposals have been made for selecting such paths which include: Equal Cost Multi-Paths ( *ECMP* ), loop-free alternative paths or multi-hop repair paths [44]. *ECMPs* are paths that do not traverse the failure, while loop-free alternative paths are established through a direct neighbor of a router adjacent to the failure. Multi-hop paths are more complex to compute. Often such paths cannot be computed/decided wholly by one router alone; for example they can be specified using a loose-hops approach or multiple routers using label based mechanisms for path discovery (label based path switching is described in more detail in the previous *MPLS − FRR* section). Often the majority of the destinations can be reached by using the first two basic path selection techniques with multi-hop path construction methods required for the remainder [44].

In fact, it is not just fast recovery that can be achieved, but traffic engineering information can be also be gleaned and paths selected accordingly to meet QoS requirements or load balancing on the links. For example, some *IGP* protocols often build up a Traffic Engineering Database. This database is typically used to optimize utilization of links inside the domain and minimize the cost of inter–domain traffic intended for an outside destination traversing its network. However, optimizing these intra-domain parameters may lead to a sub-optimal inter-domain path; e.g. shunting packets onto an inter-domain segment which is experiencing congestion. Even if the primary intra-domain path satisfies the QoS requirements for its share of the inter-domain paths, it does not guarantee that its chosen failover path would too, due to the constraints of other external domains contributing to the inter-domain path. Pre-computing such failover paths and appraising neighboring domains can yield to quick and optimal failover.

Shand and Bryant [44] highlighted several key challenges in *FRR* construction for purely *IP* networks. The first of these challenges is how a router can choose such failover paths after detecting a fault without consulting its neighbors or waiting for the protocol (e.g. *IGP* ) to converge. The second challenge is lowering complexity of computing such paths so as to not overload routers. The question is then, how to achieve an optimal tradeoff between the two?

OPEN RESEARCH ISSUES

This section overviews open research issues in the areas of multi-homing (Section V.A), modifying underlay routing mechanisms (Section V.B) and in using alternate paths in Resilient Overlay Networks (Section V.C).

*Open Research-issues with Multi-homing*

Effective multi-homing only requires that the edge network be reachable through two or more topologically diverse *ISPs* so that it can connect to the outside 'world' with reasonable assurance. Akella and Tao [45-46] considered the performance advantages using key path metrics, delay (packet round trip time), loss-rate and throughput when edge hosts are multi-homed via multiple providers and also have a choice of overlay-paths when the direct-path degrades. Such studies may be somewhat biased as they report the results from ISPs which gave best results across all destinations considered in the studies. One study [45] reported that the performance-advantage is 20-40% for delay and 15-25% for throughput, when the edge host is multi-homed via three providers. Increasing the number of providers beyond three results in marginal benefits. However, the same study also concluded that multi-homing has only limited benefits compared to when end-hosts have a choice of overlay paths between them. This is because end-to-end path diversity in the core of the Internet can only be leveraged effectively through the use of overlay networks. Another recent paper [47], stated similar results when considering the number of shared routers and physical links on alternative paths provided by multi-homing solutions.

Multi-homing provides physical redundancy while working within the *BGP* framework. However, multi-homed hosts announce their multiple routes within the *BGP* framework through different upstream-provider ISPs. Multi-homing has been blamed as one of the leading factors for the exponential increase in the size of *BGP* routing tables since 1999 [29, 48]. Multi-homing creates 'holes' in the routing table [29] because certain subsets of *IP* sub-blocks already contained within the prefix set of one of the providers of a multi-homed AS are announced again by one of the other multi-homed AS's providers for the purpose of fault tolerance. Ways of overcoming these challenges remain as open research issues.

*Open Research-issues with proposals to modify underlay routing mechanisms*

Proposals to modify underlay routing mechanisms ($BGP, IGP$) seem attractive at the outset, but pose some challenges. For example, would the path deflection decision as proposed by [30] be able to scale well enough at the individual packet level? Another core issue relates to the feasibility of implementing the proposed changes to routers to support path deflection decisions. Also, these studies solve the issue of exploitation of the path diversity of the Internet but introduce the problem of monitoring path quality, which has hampered the deployment of large overlay networks due to scalability concerns. Another area of practical concern is that redesigning underlay routing mechanisms, such as those suggested by Yang and Wetherall [30] including changes to $BGP$ [23, 49], may expose underlay routing to several security vulnerabilities. At present end systems do not exercise any control over the paths that their packets would take, which are determined solely by the network routers. Equipping end systems with the power to decide paths may open the network to compromise by an adversary or cause breach of commercial traffic transit policies between ISPs, causing conflicts over revenue.

The primary motivation of the $MPLS-TE$ solutions is not only to exploit inter-domain path diversity but also to find paths that fulfill specific QoS requirements. It is based on the premise that neighboring domains can establish trust for finding such QoS optimized paths. Since each individual domain does not have to reveal its internal structure, this trust will be weak unless there is some monetary incentive attached for it to do so. Another related issue is if the primary $LSP$ fails, each domain may have its own priority to compute restoration paths that may not be acceptable to other participating domains [42].

Such studies may also be somewhat biased as they report the results from $ISPs$ which gave best results across all destinations considered in the studies. One study [45] reported that the performance-advantage is 20-40% for delay and 15-25% for throughput, when the edge host is multi-homed via three providers. Increasing the number of providers beyond three results in marginal benefits. The same study however, also concluded that multi-homing has only limited benefits compared to when end-hosts have a choice of overlay paths between them. This is because end-to-end path diversity in the core of the Internet can only be leveraged effectively through the use of overlay networks. Another recent paper [47], stated similar results when considering the number of shared routers and physical links on alternative paths provided by multi-homing solutions. There is clear scope for further research in this field.

*Open Research-issues with Resilient Overlay-Networks*

Qiu et al [50] noted that selfish-routing using RONs can harm traffic-engineering goals. Overlays choose paths which are longer than direct Internet paths and may prefer certain links more than others. This increases network load and increases congestion on some links as investigated by [50]. Recent debates [51-52] on coexistence of multiple overlays and their co-existence with the (non overlay) Internet traffic have aroused suspicions about the effectiveness of overlays in the long term. It is well understood now that overlay routing networks can improve performance by leveraging the inherent path redundancy in the Internet. However, they do so by transferring the traffic from one subset of paths to another. Keralapura et al [52] conjectured that multiple overlays performing the same function using their own greedy and selfish routing metrics in selection of overlay paths could introduce *race* conditions leading to unwanted routing oscillations (Figure 10). They found that the probability with which two overlay networks can get synchronized increases if the multiple interacting overlays are aggressive i.e. have short path probing intervals or path outage detection times close to each other. This can happen if the overlay hosts of multiple overlay networks are situated close to each other, leading to similar path round trip times used for probe timeouts, an indicator of path failure. The more dissimilar the overlay networks are in terms of locality of nodes and path probing parameters, the lower the probability of routing oscillations [52].

Another issue is that all major research into overlay-network behavior revolves around analytic evaluations using simulated topologies [53-55] or limited deployments of a few overlay test beds [5, 56] and a few selected large $ISPs$, e.g. Abilene and Sprint [34, 57] due to the difficulties in practical deployment. A majority of the work using simulated topologies uses the hierarchical power-law model [36] to build the underlay (and overlay) topology. However, some recent works [58-59] give substantial evidence that such static power-law models may not capture the Internet topology accurately enough because the Internet evolution is dynamic process shaped by a several interconnected variables. Thus the results derived from them could potentially be inaccurate and misleading.

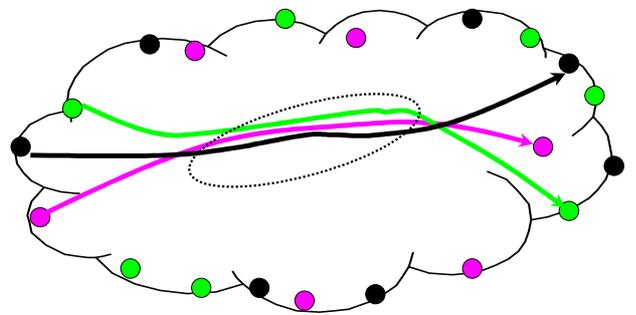

Figure 10. Contention for same set of physical links. Three overlay networks decide to use same set of physical links to improve QoS on end-to-end paths increasing network load (congestion) on links and also towards possible oscillations in quest for better paths.

A final open research issue is that of constructing overlay networks offering high inter-domain path diversity. In*tra*-domain path diversity is often not a concern as an *IGP* can quickly re-route packets using better paths. In*ter*-domain paths, on the other hand, suffer from delayed convergence after faults as explained earlier. Overlay hosts often cannot control their location. It is a challenge to place overlay peers across inter-domain regions so as to provide maximum path diversity. Han et al. [47] reported that even if overlay nodes were located in topologically diverse ISPs, their paths to a destination often traversed the same links/routers. They also noted that overlay peers must not be in ISPs that maintain peering relationships with each other. *BGP* often hides such information. Provider, customer peering relationships can only be inferred from the study of *BGP* dumps.

CONCLUSIONS AND PROPOSALS FOR FUTURE DIRECTIONS OF RESEARCH

BGP can suffer from delayed convergence after failure, and Internet flows seeking QoS guarantees may seek alternate paths quickly to mask such failures. In this paper, we surveyed several contemporary techniques for discovering and using alternate paths, which the research community has come up with in recent years. What follows are our observations, based on this survey, for the future of research for development of overlay networks, modifications to underlay routing mechanisms and multi-homing solutions.

*Resilient Overlay Networks: Research needs more deployment to back simulation results*

Resilient Overlay Networks seek advantage from the physical redundancy in the core and edge of the Internet to discover end-to-end paths. They offer large design freedom to compute alternate paths dynamically but correspondingly require extra overheads for path performance information, or intelligent topology-aware design to predict good alternate paths if such information is not available.

Current research aimed at reducing path monitoring overheads by leveraging topological knowledge seem to be the most promising area of research at the moment [33-34, 57, 60] but unfortunately the performance benefits they claim to have only been demonstrated through limited deployment. These issues need to be addressed in detail using real heterogeneous overlay deployments in the Internet with limited topological knowledge [18] to fully ascertain their benefits beyond the stated claims. Similar arguments apply to overt criticisms in the research community directed against selfish routing by overlay networks [51-52].

On the other hand, while the research [21, 61-62] based on exploiting Internet path redundancy using the framework of existing Content Distribution Networks (*CDNs*) does not have astounding statistics that surprise the research community, it does present a more scalable (and plausible) avenue for the evolutionary infrastructure development. Such *CDNs* are already deployed and have been around for some time now and scalable methods for determining the locality of peers and topology maintenance have been developed. We think that these studies warrant more attention from the research community.

*Multi-homing will stay but its implementation will change*

Multi-homing will continue to be used to provide path redundancy for stub ASes. We doubt much will happen to this stagnated area for exploration of Internet path diversity. However, its implementation may change due to the one big cause of the concern: its contribution to inflation of *BGP* routing tables [29, 48]. Multi-homed ASes may experience a shift from a static binding between end point identifiers (IP addresses) and locations (multiple routes), to a more flexible architecture [63] that binds intermediate location identifiers (LIDs) to end point identifiers (EIDs) to reduce the load on routing tables. Even with these implementation changes, the potential benefits for multi-homing to explore path diversity in the Internet continue to be limited to the edge (and not the core) of the Internet, and are thus aimed more towards path outages rather than QoS enhanced paths.

*Incorporating reverse engineering to underlay routing mechanisms may leave them vulnerable to security issues and cause revenue conflicts between ISPs*

Proposals to re-engineer underlay routing mechanisms were primarily aimed at reducing underlay convergence time after path failures [2] and in providing end users with the flexibility of being able to choose their own end to end routes look promising. However, a disadvantage is that their scalability may be challenged as several of these proposals entail per packet or per session overhead.

The second issue is related to security: allowing end users to influence path selection also equips them to launch various malicious attacks aimed at *ISPs*.

The third issue is related to the violation of commercial traffic policies agreed upon by *ISPs* where routes are dictated more by financial agreements rather than stringent QoS requirements on a per user basis. Proposals such as *MIRO*, *NIRA* [31, 64] may cause revenue conflicts between *ISPs* that are not a problem at the moment because *BGP* traditionally allows for a level playing field for all network players. Nevertheless, these are still in preliminary stage of development and it will take some time and community-effort (and trust!) to deploy.

A fourth issue is again regarding the credibility of simulation platforms that aim to model $BGP$; e.g. $C$-$BGP$[1], $SSFNet$[2]. Such simulators limit network sizes due to scalability issues resulting from memory and processing power required. For example, $C-BGP$ only simulates the $BGP$ decision process but does not incorporate other variables such as $BGP$ timer implementations to keep the simulation platform scalable. Hence, it cannot be effectively used to investigate the performance of $BGP$ convergence in all desired scenarios and dimensions.

While alternate paths are sure to play an increasing role in future Internet routing, there remains considerable scope for more research in this area.

<div style="text-align: center;">REFERENCES:</div>

---

[1] C-BGP. Université Catholique de Louvain, Belgium. see http://cbgp.info.ucl.ac.be/

[2] SSFNet. Scalable Simulation Framework. See http://www.ssfnet.com/

Sameer Qazi received the B.E. degree in Electrical Engineering from the National University of Science and Technology (NUST), Pakistan, in 2001 and the M.Eng.Sc. degree in Telecommunications from the University of New South Wales, Australia, in 2004. He is currently an Assistant Professor in National University of Sciences and Technology, Pakistan. Previously, he has worked in WorldCALL Broadband limited as Fiber Optics Engineer in Pakistan (2001–2003). His current research interests include techniques for improving Reliability of Routing Overlays, Network Optimization, QoS provisioning, Peer to Peer systems and Geo-based location services for Internet hosts. He is a member of IEEE and ACM and a lifetime member of the Pakistan Engineering Council. He is a PhD approved supervisor in Pakistan and his name appears in the 30[th] (Pearl Anniversary) Edition of Marquis Whos Who in the World.

Tim Moors is a Senior Lecturer in the School of Electrical Engineering and Telecommunications at the University of New South Wales, in Sydney, Australia. He researches network reliability, transport protocols, and wireless LAN MAC protocols. Previously, he was with the Center for Advanced Technology in Telecommunications at Polytechnic University in New York, and prior to that, with the Communications Division of the Australian Defence Science and Technology Organisation. He received his Ph.D. and B.Eng.(Hons) degrees from universities in Western Australia (Curtin and UWA).